# Using Dust as Probes to Determine Sheath Extent and Structure

ANGELA DOUGLASS[1], VICTOR LAND[2], KE QIAO[3],
LORIN MATTHEWS[3], AND TRUELL HYDE[3]

[1]Ouachita Baptist University, 410 Ouachita St., Arkadelphia, Arkansas 71998, USA
[2]Eindhoven University of Technology, 5600 MB Eindhoven, the Netherlands
[3]Center for Astrophysics, Space Physics, and Engineering Research, Baylor University,
Waco, Texas 76798, USA

Two in-situ experimental methods are presented in which dust particles are used to determine the extent of the sheath and gain information about the time-averaged electric force profile within a RF plasma sheath. These methods are advantageous because they are not only simple and quick to carry out, but they also can be performed using standard dusty plasma experimental equipment. In the first method, dust particles are tracked as they fall through the plasma toward the lower electrode. These trajectories are then used to determine the electric force on the particle as a function of height as well as the extent of the sheath. In the second method, dust particle levitation height is measured across a wide range of RF voltages. Similarities were observed between the two experiments, but in order to understand the underlying physics behind these observations, the same conditions were replicated using a self-consistent fluid model. Through comparison of the fluid model and experimental results, it is shown that the particles exhibiting a levitation height that is independent of RF voltage indicate the sheath edge - the boundary between the quasineutral bulk plasma and the sheath. Therefore, both of these simple and inexpensive, yet effective, methods can be applied across a wide range of experimental parameters in any ground-based RF plasma chamber to gain useful information regarding the sheath, which is needed for interpretation of dusty plasma experiments.

## 1. Introduction

When a plasma is created within a confined region, such as within a vacuum chamber, the plasma-facing surfaces of the chamber obtain a negative charge as the high-mobility electrons within the plasma collide with the surface. A boundary layer, called a *sheath*, immediately forms along each of these plasma-facing surfaces to shield the interior, quasineutral region of the plasma, called the bulk, from the charged surfaces. Thus, a potential difference proportional to the electron temperature is present across the sheath. The thickness of the sheath is between a few to ten times the electron Debye length, which, for low-temperature, radio frequency (RF) plasmas (such as those discussed in this paper), corresponds to a distance of millimeters to a centimeter. The time-averaged electric field generated within the sheath is directed toward the boundary surface and, within low-temperature RF plasmas, generally reaches magnitudes of several thousands of Volts per meter (Bonitz 2010; Land 2013). (Only time-averaged values and effects within an RF plasma are considered throughout the remainder of this paper as the dynamical time scales for quantities of interest are much longer than a single RF cycle.)

Small, usually micron-sized particles, may be added to the plasma to create a complex, or dusty plasma (Bonitz 2010). These particles also charge negatively due to the high mobility of the electrons within the plasma. In Earth-based experiments, the dust particles levitate within the sheath above the lower electrode where the electric and gravitational forces on the particle are equal. Therefore, the levitation height is dependent on the electric field profile within the sheath as well as the charge-to-mass ratio of the particles. Both the electric field profile and the dust charge are shaped by the plasma parameters within the sheath. If these parameters - such as ion and electron densities and temperatures - were readily available, then interpretation of most Earth-based dusty plasma experiments would be much simpler. However, direct measurements of plasma parameters are difficult to obtain. Laser-induced fluorescence (LIF) has been used for this purpose, but it requires expensive equipment, complicated diagnostic techniques, and is possible for only a small range of parameters (Vladimirov 2005). Langmuir probes inserted into a plasma have also been used to determine the plasma parameters, but the insertion of a probe within the sheath region of the plasma alters the plasma parameters and skews the results (Tomme 2000; Sheehan 2011). While knowledge of these plasma parameters is not necessary to determine levitation height of the particles, as that can be experimentally observed, it is important to know the electric field, electric force, and plasma environment experienced by a dust particle in order to analyze any dust particle behavior and interactions in dusty plasma experiments.

Theoretical models have been employed to determine the plasma parameters in the sheath and, therefore, the electric field profile within the sheath. Although many models have been proposed regarding the sheath structure and extent, the solutions are still heavily debated (Ivlev 2000; Zafiu 2001; Bohm 1949; Brinkmann 2007). In addition, they require measurements of the temperature and number density of ions and electrons within the plasma, which are difficult to experimentally obtain, making it difficult to directly apply the theoretical models to an experiment. Recent experimental results (Douglass 2012), when compared to several leading theories (Bohm 1949; Brinkmann 2007), showed that although theoretical models may come close to describing the nature of the sheath, aspects of these theories are only valid within specific regions of the sheath, generally near the center of the sheath (Douglass 2012). Deviations are found between the models and experiment near the point where the sheath and bulk meet, called the sheath edge. At this point the plasma parameters become increasingly complex making accurate modeling of this region difficult. Dusty plasma experiments, for which the particles are sufficiently far away from the sheath edge, could reasonably use theoretical models to calculate the plasma environment surrounding the particles, but these models are not valid for experiments in which dust particles are located near the sheath edge. Therefore, determination of the location of the sheath edge becomes an important aspect in dusty plasma experiments.

The methods described in this tutorial use the dust particles themselves as probes to determine the electric force as a function of height above the lower electrode. The small size of the dust grain allows for minimal perturbation of the plasma parameters within the sheath, providing more accurate information regarding its extent and structure. Other experimental methods have been employed to investigate the plasma sheath, but often require additional equipment (Beckers 2011; Maurer 2011). The following methods are not only simple to carry out, but use only standard dusty plasma experimental equipment such as dust, cameras, and lasers already present in dusty plasma laboratories.

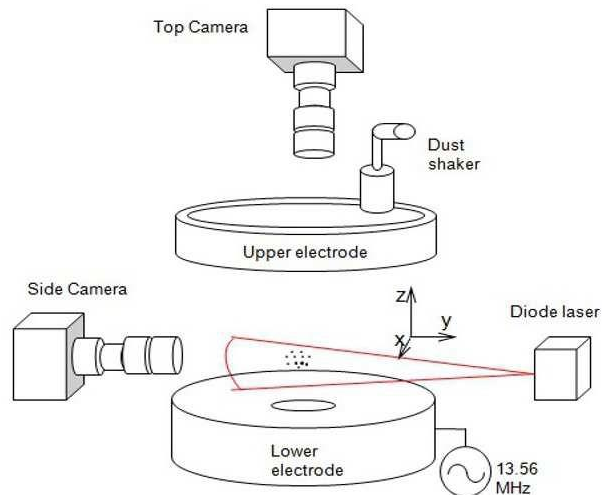

Figure 1. The CASPER experimental RF plasma chamber used in the following experiments.

In Section 2, an explanation of two experimental techniques that can be used to determine the electric force as a function of height above the lower electrode and the location of the sheath edge will be presented. A description of the interpretation and application of these techniques is given in Section 3. A final summary is given in Section 4.

## 2. Experimental Method

Both of the following methods use the dynamics of the dust particles themselves to determine the sheath structure and extent and were employed in the Center for Astrophysics, Space Physics and Engineering Research (CASPER) Gaseous Electronics Conference (GEC) cell at Baylor University (Hargis 1994; Land 2009). A schematic of the cell is shown in Figure 1. The cell contains a powered (capacitively coupled), lower electrode driven at 13.56 MHz. Experiments were performed using melamine-formaldehyde (MF) particles with a mass density of 1.510 g/cm$^3$ in argon plasma at a pressure of 20 Pa with RF voltage amplitudes ranging between 22 V and 66 V (corresponding to a RF power of 0.9 W to 4.3 W) and an external power supply was used to maintain a DC bias of -5 V on the lower electrode, but the techniques are applicable to any GEC RF plasma cell operating across all normative pressures and powers.

### 2.1. Experimental Determination of the Electric Force Profile

The electric force as a function of height above the lower electrode can be determined by tracking the dust particles as they fall through the plasma toward the lower electrode. First, a viewing laser with a vertical fan is required to observe particles along their entire vertical trajectory. A Coherent LASIRIS Laser of wavelength 685 nm was used in this experiment, but any laser that provides illumination of the particles without perturbing them is sufficient. In addition, a high speed camera is required to obtain multiple images of the particle as it descends. These experiments used a Photron 1024 high-speed camera operating at 1000 frames per second (Douglass 2012).

The vertical motion of the particle is given by
$$m_D \ddot{z} = F_E(z) - m_D g - m_D \beta \dot{z} \qquad (2.1)$$
where β is the Epstein drag coefficient (Epstein 1924), $m_D$ is the dust particle mass, and z is the height of the particle above the lower electrode. β can be calculated from the equation
$$\beta = \frac{8}{\pi} \frac{P}{\rho r_D v_{th,n}} \qquad (2.2)$$
where P is the gas pressure, ρ and $r_D$ the mass density and radius of the dust particle, respectively, and $v_{th,n}$ is the thermal velocity of the neutral gas. Although the particle also experiences a force due to ion drag, this force is not included in the equation of motion given in Equation 2.1 since it has been estimated to be significantly smaller than the included forces (less than 3% of the gravitation force) using the fluid model described in Section 3. This is in agreement with similar experiments which have found the ion drag force on particles with similar sizes to those used here to be negligible with respect to the electrostatic and gravitational forces (Beckers 2011; Basner 2009).

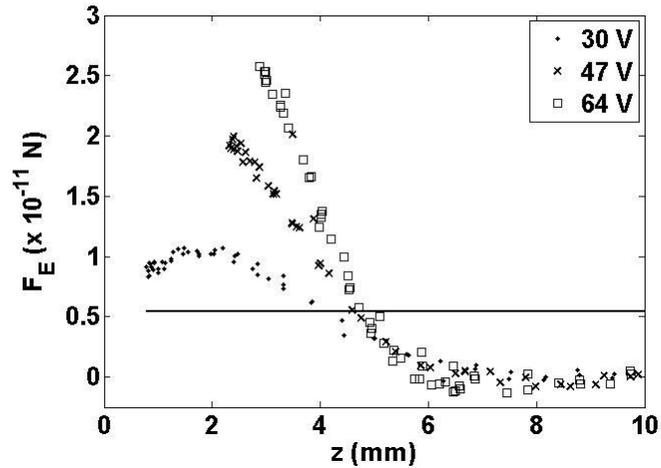

Figure 2. The experimentally determined electric force profile at a pressure of 20 Pa for various RF voltages. The horizontal line indicates the magnitude of the gravitational force acting on an 8.89 μm MF particle (Douglass 2012).

By locating the particle's position in each frame, a determination of the particle's vertical velocity, $\dot{z}$, and vertical acceleration, $\ddot{z}$ can be made. Substituting these values along with the mass of the particle into Equation 2.1, the electric force as a function of height can then be determined. Figure 2 shows a representative graph of the electric force as a function of height above the lower electrode for an 8.89 μm-diameter particle obtained using this method (Douglass 2012).

It is important to note that the magnitude of the electric force will vary depending on particle size, as the dust particle charge varies linearly with particle radius. However, the overall shape of the electric force profile will remain the same since the electric field is a property of the sheath itself and not the particle.

*2.2. Determination of the Sheath Edge through Levitation Height*

Since the dust particles levitate at a height where the gravitational and electric forces on the particles are equal, the following method uses the levitation height of the dust particles to determine sheath edge information as well as information about the vertical electric field structure.

First, a small number of dust particles (10-50) of a single size are added to the plasma and allowed to form a single, horizontal layer at equilibrium. Side-view images of the layer are then recorded to obtain the levitation height of the particles above the lower electrode. Next, levitation height of the particles should be recorded across a range of RF voltages. This procedure should be completed multiple times, using dust particles with a different diameter each time in order to obtain a graph of the levitation height versus RF voltage such as that shown in Figure 3 (Douglass 2012).

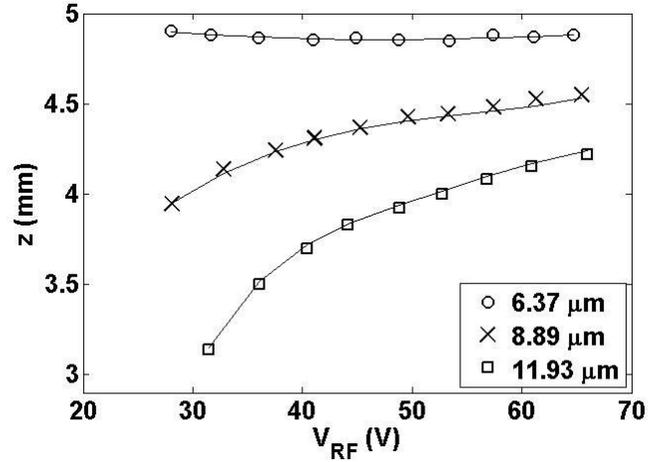

Figure 3. Experimentally measured levitation heights for layers consisting of different-sized particles as a function of RF voltage. The levitation height of each particle size was measured in separate experiments with neutral gas pressure of 20 Pa. Lines are added to guide the eye (Douglass 2012).

## 3. Discussion

Interesting trends are observed in each of the previous figures. Notice, in Figure 2, for z > 5mm, the electric force profiles for all RF voltages tested coincide. This indicates that the electric force is independent of the RF voltage in this region. In Figure 3, it is observed that the levitation height of large particles increases with increasing RF voltage and appears to approach a limiting value, while the levitation height of the smallest particles remained relatively constant. In order to understand the physics behind these behaviors and the underlying sheath structure, a fluid model of the plasma in a GEC RF reference cell (the specifics of which can be found in Land (2010)) was employed.

The fluid model used solves the plasma parameters self-consistently for given operating conditions within the 2D geometry of the CASPER cell, providing a reliable method to check experimental results against theory. When applied to the operating conditions used in the previous experiments, the fluid model generated similar electric force profiles and levitation height dependence given similar operating conditions as the experiments. An example of the electric field profile obtained from the fluid model is shown in Figure 4.

The plasma parameters found through the fluid model, $n_e(z)$, $n_i(z)$, $T_e(z)$, were used to calculate the currents to the dust particles and thus the dust surface potential, $\Phi_D$, which was then used to calculate the charge on a particle through use of the capacitor model ($Q_D = 4\pi\epsilon_0 r_D \Phi_D$). The electric force profile was then obtained using the formula $F_E(z) = Q_D(z)E(z)$, where again, the electric field, E(z) is found from the fluid model. The electric force profiles for 8.89 μm and 16 μm particles is shown in Figure 5. Notice that the electric force profile obtained from the fluid model exhibits the same overall shape as that found experimentally (Figure 2). In addition, the fluid model results showed that the electric field, and the electric force, when determined for a range of RF voltages, converge to a single value just as observed in Figure 2 (Douglass 2012). The explanation for this convergence was determined through investigation of the ion and electron densities obtained in the fluid model. Figure 6 shows the ratio of the ion density to electron density, $\alpha$, as a function of height for various RF voltages. The horizontal line indicates where the electron and ion densities are equal, locating the point where quasineutrality is obtained. Notice that the height at which quasineutrality is obtained appears independent of RF voltage. This height is the same height that the electric field profiles converge in Figure 4 and the electric force profiles converge in Figure 5. Thus, the convergence of the electric field or electric force signifies the location where quasineutrality is obtained - i.e., the sheath edge.

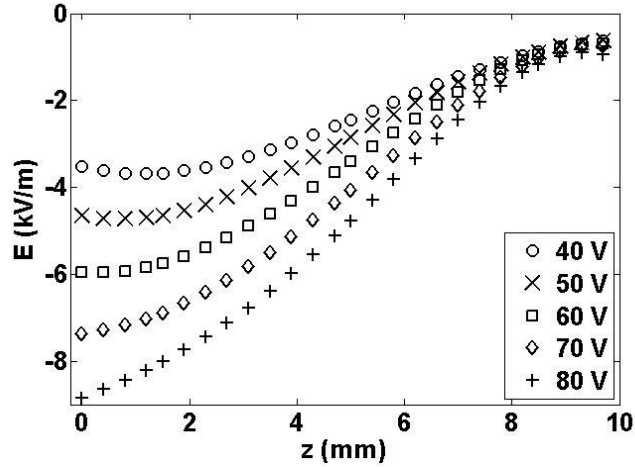

Figure 4. The electric field profile for various RF voltages at 20 Pa obtained from the fluid model (Douglass 2012).

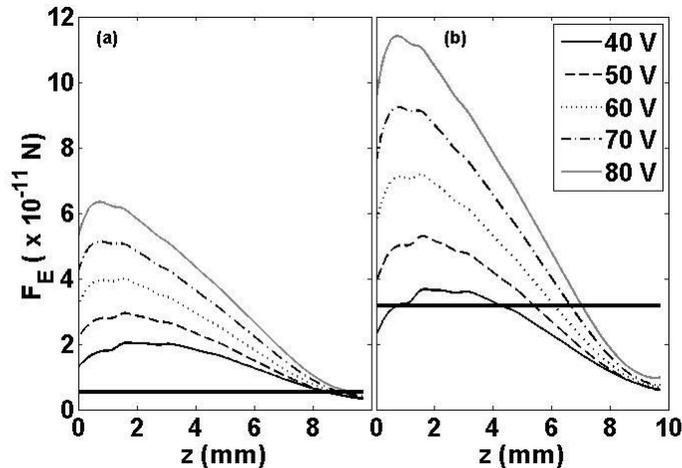

Figure 5. The electric force profile for various RF voltages at 20 Pa for (a) 8.89 μm and (b) 16 μm diameter particles obtained from the fluid model (Douglass 2012).

While it is fairly simple to determine the vertical electric force profile experimentally, determination of the electric field, E(z), or dust charge, $Q_D(z)$, independently are difficult to obtain. However, the results of the fluid model show that the dust surface potential, and therefore the charge on a dust particle, remains relatively constant within the sheath as can be seen in Figure 7. Therefore, the structure of the electric force profile is mainly determined by the electric field. So, in looking back at the experimental results (Figure 2), the convergence of the electric force profiles, which occurs at z ≈ 5mm, indicates the height above the lower electrode at which the electric field profiles converge signifying that quasineutrality is obtained, identifying the sheath edge. It is important to note that the sheath edge location as determined from the fluid model (z = 9 mm from Figure 5) and that located from the experiment (z = 5 mm from Figure 2) are not at the same height. While the fluid model was able to provide qualitatively similar results which can be used to interpret the experiments, some quantitative differences were observed, most likely due to approximations and limitation of the fluid model (see Douglass (2012) for a description of these limitations).

In addition, the experimental electric force profile can be used to determine the lower levitation limit of a particle. An unstable region exists near the lower electrode which is characterized by a positive gradient in the electric force (Ivlev 2000). Therefore, the lower levitation limit for a particle corresponds to the height at which the electric force profile obtains a maximum value near the lower electrode. This fact has recently been used to manipulate the number of particles within a crystal or vertically extended dust chain (Kong 2011).

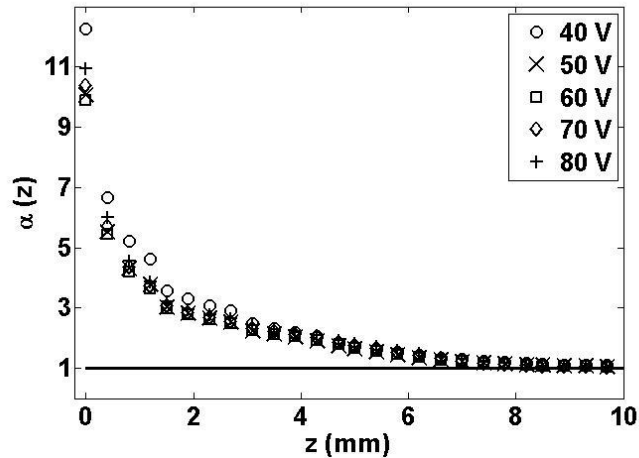

Figure 6. The ratio of the ion density to the electron density, $\alpha(z)$, in the sheath for various driving potentials, $V_{RF}$, at 20 Pa obtained from the fluid model. The horizontal line indicates where the electron and ion densities are equal (Douglass 2012).

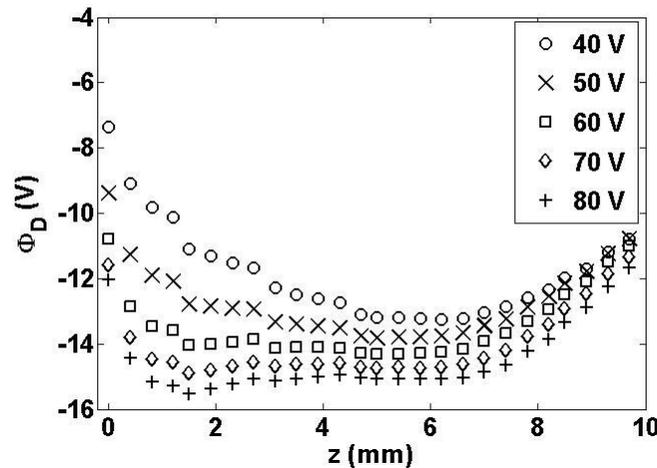

Figure 7. The dust surface potential for various driving potentials, $V_{RF}$, at 20 Pa obtained from the fluid model (Douglass 2012).

Upon further investigation it can also be seen that the electric force profiles, such as Figure 2, can be used to interpret the levitation height trends shown in Figure 3. The solid, horizontal line in Figure 2 indicates the gravitational force on the particle. The levitation height of a particle at a certain RF voltage can then be read directly from Figure 2 by simply locating the height at which the horizontal, gravitational force line intersects with the electric force curve. From Figure 2, it can be seen that the levitation height for 8.89 μm particles increases as the RF voltage increases. This increase in height is nonlinear, with larger increases in height at low RF voltages and smaller increases at higher RF voltages. This correlates with the trend shown in Figure 3 for the 8.89 μm particles. Notice, though, that the levitation height of the 6.37 μm particles, as shown in Figure 3, does not increase as RF voltage increases, but rather remains nearly constant. From this it is concluded that the 6.37 μm particles levitate at a height where the electric force (and also electric field) profiles converge to a constant value, which was previously determined to be the sheath edge. Therefore, the sheath edge location can be determined for various plasma parameters by locating the height at which the change in levitation height with change in RF voltage is approximately zero, $dz/dV_{RF} \approx 0$, across a wide range of RF voltages.

It is important to note that this method of locating the sheath edge is limited by the resolution of the experiment. An infinite number of different sizes of particles as well as an extremely high resolution camera would be required to precisely determine the location of the sheath edge. Nonetheless, this simple method provides a good approximation of the location of the sheath edge. In addition, it is important to note that a small, ambipolar electric field located at and above the sheath edge may be sufficient to levitate small particles outside the sheath within the plasma bulk. These particles could exhibit the same trends as particles located at the sheath edge. Therefore, if multiple particle sizes display levitation heights in which $dz/dV_{RF} \approx 0$, across a wide range of RF voltages, only the largest-diameter particles exhibiting this behavior should be considered to levitate at the sheath edge. An exact particle diameter cannot be given, since this will depend on the experimental parameters.

## 4. Conclusions

Two simple, inexpensive experimental methods that can be used to determine the time-averaged extent of the sheath and its structure through use of the dust particles as probes have been presented. The first method employs the use of a high speed camera to track a dust particle as it falls toward the lower electrode in a GEC RF plasma. The location of the particle in each image was then used in Equation 2.1 to calculate the electric force on the particle as a function of height. This alone provides necessary information in the interpretation of dusty plasma experiments since it quantifies a force experienced by a dust particle. Additionally, through use of a fluid model, it was shown that the height at which the electric force profiles for various RF voltages converge indicates the location where quasineutrality within the plasma is obtained - i.e., the sheath edge. Since the fluid model results qualitatively match experimental results (for example Figure 2 and Figure 5 show the same trend), the conclusion that the convergence of the electric force profiles indicates the sheath edge can then be applied to the experimental results. So, the sheath edge can be located through an experiment alone from graphs such as Figure 2 through determination of the height where electric force profiles for various RF voltages begin to coincide. This allows one to determine if the dust particles used in other experiments under the same conditions are near the sheath edge and provides insight into whether or not leading theoretical sheath models can be applied. In addition, the lower levitation limit of the dust particle can be determined from this method, which is useful information in experiments such as those involving dust particle chains (Kong 2014).

In the second method, the levitation of a small number of dust particles (10-50) across a wide range of RF voltages was obtained. Since levitation height is dependent on the electric force on the dust particle, graphs such as Figure 3 can be used to gain insight into the electric force as a function of particle height above the lower electrode. It was shown that the largest-diameter particles that display a nearly constant levitation height across a range of RF voltages can be used to locate the sheath edge. This method provides a good approximation of the location of the sheath edge, but is limited by the dust particle sizes available and the camera resolution.

As previously stated, both of these methods are applicable across a wide range of experimental parameters and in any ground-based RF plasma chamber. Any experiment exhibiting either the convergence of the electric force profile or the constant levitation height of a particle with RF voltage as shown here can be used to determine the sheath edge. Since these experimental methods only require standard equipment that is already employed in dusty plasma experiments, they are quick, simple, and inexpensive to implement, yet provide essential information for interpretation of dusty plasma experiments. In the future, refinements to these techniques may allow for them to be applied to magnetic or microgravity dusty plasma systems as well.

This material is based upon work supported by the National Science Foundation under Grant No. 0847127 and by the Texas Space Grant Consortium through a graduate fellowship.